\documentstyle[12pt]{article}
\catcode`\@=11
\begin{document} \openup6pt

\title{INFLATON FIELD AND PRIMORDIAL BLACKHOLE}

\author{B. C. Paul*\thanks{ e-mail : bcpaul@nbu.ernet.in } and S. Chakraborty**\thanks{e-mail : subenoy@juphys.ernet.in}\\
   Inter-University Centre for Astronomy and Astrophysics \\
P.O. Box. : 4, Ganeshkhind, Pune-411007, INDIA \\
** Department of Physics, North Bengal University, \\
   Siliguri, Dist. Darjeeling, Pin : 734 430, INDIA \\
*** Department of Mathematics, Jadavpur University, \\
   Kolkata - 700 032, INDIA }

\date{}

\maketitle

\begin{abstract}

Primordial black hole formation has been studied using an inflaton field with a variable cosmological term as the potential.
 \\

PACS number(s) : 04.20.Jb, 04.60.+n, 98.80.Hw
\end{abstract}

\vspace{1.0in}

\pagebreak

\section{INTRODUCTION : }

In astrophysics, it is known that the  black holes are the ultimate corpse of a collapsing massive star when  the  mass of a  given star exceeds about twice that of 
the solar mass. It is  probable in cosmology that other kind of black holes having mass smaller than the solar mass may have produced. These are lighter so are hotter and important in the context of determination of  Hawking radiation.$^{1}$  The later types of black holes are popularly known as primordial black holes (henceforth, PBH).  Such kind of blackholes are likely to be formed during quantum fluctuation of the matter fields in the early universe during inflation. The inflationary scenario of the early universe are derived out of an inflaton field guided by a potential during its slow rolling approximation. The condition that PBH may produce requires a cosmological constant was obtained by this slow roll approximation as the potential can be treated as a constant (flat potential) during the short inflationary regime.$^{2}$   Although the chance of getting PBH's are small today may have been more  in the very early universe. They might have played a crucial role in the structure formation of the late universe.$^{3}$  Recently, Meng 
{\it et al.}$^{4}$  studied the creation of PBH from inflaton field starting from the static spherically symmetrical de Sitter black hole spacetimes with charge and mass. They examined the PBH properties, for both charged and uncharged  cases and obtained a relationship between the primordial black hole pair formation and the energy transfer of the inflaton field.

In this work we consider a special form of Reissner-Nordstr$\ddot{o}$m-de Sitter model with metric ansatz$^{5}$  given by 
\begin{equation}
ds^{2} = - A(r, t)dt^{2} + A(r, t)^{-1} dr^{2} + r^{2} d\Omega_{2}^{2}
\end{equation}
with 
\[
A(r, t) = \left( 1 - \frac{a}{r} \right)^{2} \left[ 1 - \frac{\Lambda}{3} \left( r^{2} + 2 a r + 3 a^{2} \right) \right]
\]
where $a$ is a constant and $\Lambda$ is the cosmological  term. The
mass and charge of the corresponding black hole solution are given by
\begin{equation}
{\bf M} = a \left( 1 - \frac{2}{3} \Lambda a^{2} \right),
\end{equation}
\begin{equation}
{\bf Q}^{2} = a^{2} \left( 1 -  \Lambda a^{2} \right).
\end{equation}
Here, the cosmological parameter $ \Lambda $,  is taken to be a function of the homogeneous scalar field $\phi(t)$ which causes inflation. We choose for $\Lambda$ the the form of potential used by Peebles and Vilenkin$^{6}$  as 
\begin{equation}
\Lambda (\phi) = b \left( \phi^{4} + \phi_{o}^{4} \right).
\end{equation}
It  is to be noted that in the pre-inflationary era, $\phi$ dominates the dynamics and hence $\Lambda$ is considered as a variable quantity. But as the inflationary expansions proceeds, $\phi$ gradually  becomes insignificant and at present we may take $\phi$ to be a constant. 

Further, in order to solve the flatness problem during the inflation we must have
\begin{equation}
 \frac{1}{2}  \left(  \frac{\Lambda'}{\Lambda}  \right)^{2} = \frac{ 8 \phi^{6}}{(\phi_{o}^{4} + \phi^{4})^{2}} << 1
\end{equation}
and 
\begin{equation}
\frac{\Lambda ''}{\Lambda} = \frac{ 12 \phi^{2}}{(\phi_{o}^{4} + \phi^{4})} << 1.
\end{equation}
If we introduce, $\rho = r/{\bf M} $, $z = \Lambda {\bf M}^{2}$, then we get from $ A(r, t) = 0$
\begin{equation}
z = \frac{3 \left( 1 + \frac{2 {\bf Q}^{2}}{a^{2}} \right)^{2}}{\left[
\rho^{2} \left( 1 + \frac{2 {\bf Q}^{2}}{a^{2}} \right)^{2} + 6 \rho 
\left( 1 + \frac{2 {\bf Q}^{2}}{a^{2}} \right) + 27 \right]}.
\end{equation}
We note that $  A(r, t) = 0$ has  two non-degenarate and a degenerate roots so \begin{equation}
\frac{dz}{d\rho} = 0
\end{equation}
gives us a unique root 
\begin{equation}
\rho  = - 3 /\left(1 + 2 {\bf Q}^{2}/a^{2} \right) = \rho_{-}
\end{equation}
with $z(\rho_{-}) = \frac{1}{6} \left(1 + 2 {\bf Q}^{2}/a^{2} \right)^{2}$.

Thus we have the following two situations

(i) if  $z(\rho_{-}) = \Lambda {\bf M}^{2}$, then there is only one horizon

(ii) if  $z(\rho_{-}) > \Lambda {\bf M}^{2}$, then no horizon exists.

The first situation is known as the critical solution and can be written in explicit form
\begin{equation}
b \left( \phi_{o}^{4} + \phi^{4} \right) {\bf M}^{2} = \frac{1}{6} \left[
1 +   \frac{2 {\bf Q}^{2}}{9 {\bf M}^{2}} \left( 1 + \frac{2 {\bf Q}^{2}}{a^{2}} \right)^{2}  \right]^{2}
\end{equation}
provided 
\begin{equation}
8 {\bf Q}^{2} < 9 {\bf M}^{2}. 
\end{equation}
However for an uncharged case we have
\begin{equation}
b \left( \phi_{o}^{4} + \phi^{4} \right) {\bf M}^{2} = \frac{1}{6} \end{equation}

To conclude we obtain the criterion for a primordial black hole formation using a variable cosmological constant as suggested be Peebles and Vilenkin as the potential for the inflaton field.

{\it Acknowledgements  : }

The authors (BCP and SC)  like to thank 
the Inter-University University Centre for Astronomy
and Astrophysics ( IUCAA ), Pune for warm hospitality and providing facilities to complete this work during a visit. 

\pagebreak

\end{document}